\documentclass[aip,apl,amsmath,amssymb,reprint]{revtex4-1}
\usepackage{graphicx}
\usepackage{dcolumn}
\usepackage{bm}
\usepackage{color}

\begin{document}

\title{Resonantly tunable second harmonic generation from lithium niobate metasurfaces}

\author{Junjun Ma}
\affiliation{The Key Laboratory of Weak-Light Nonlinear Photonics, Ministry of Education, School of Physics and TEDA Applied Physics Institute, Nankai University, Tianjin 300071, P.R. China}

\author{Fei Xie}
\affiliation{The Key Laboratory of Weak-Light Nonlinear Photonics, Ministry of Education, School of Physics and TEDA Applied Physics Institute, Nankai University, Tianjin 300071, P.R. China}

\author{Weijin Chen}
\affiliation{School of Optical and Electronic Information, Huazhong University of Science and Technology, Wuhan, Hubei 430074, P.R.China}

\author{Jiaxin Chen}
\affiliation{The Key Laboratory of Weak-Light Nonlinear Photonics, Ministry of Education, School of Physics and TEDA Applied Physics Institute, Nankai University, Tianjin 300071, P.R. China}

\author{Wei Wu}
\affiliation{The Key Laboratory of Weak-Light Nonlinear Photonics, Ministry of Education, School of Physics and TEDA Applied Physics Institute, Nankai University, Tianjin 300071, P.R. China}

\author{Wei Liu}
\affiliation{College for Advanced Interdisciplinary Studies, National University of Defense Technology, Changsha, Hunan 410073, P.R.China}

\author{Yuntian Chen}
\affiliation{School of Optical and Electronic Information, Huazhong University of Science and Technology, Wuhan, Hubei 430074, P.R.China}

\author{Wei Cai}
\affiliation{The Key Laboratory of Weak-Light Nonlinear Photonics, Ministry of Education, School of Physics and TEDA Applied Physics Institute, Nankai University, Tianjin 300071, P.R. China}

\author{Mengxin Ren}
\email{ren$_$mengxin@nankai.edu.cn}
\affiliation{The Key Laboratory of Weak-Light Nonlinear Photonics, Ministry of Education, School of Physics and TEDA Applied Physics Institute, Nankai University, Tianjin 300071, P.R. China}
\affiliation{Collaborative Innovation Center of Extreme Optics, Shanxi University, Taiyuan, Shanxi 030006, P.R. China}

\author{Jingjun Xu}
\email{jjxu@nankai.edu.cn}
\affiliation{The Key Laboratory of Weak-Light Nonlinear Photonics, Ministry of Education, School of Physics and TEDA Applied Physics Institute, Nankai University, Tianjin 300071, P.R. China}

\date{\today}

\begin{abstract}
Second harmonic generation (SHG) is a coherent nonlinear phenomenon that plays an important role in laser color conversion. Lithium niobate (LN), which features both a large band gap and outstanding second-order nonlinearities, acts as an important optical material for nonlinear frequency conversion covering a wide spectral range from ultraviolet to mid-infrared. Here we experimentally demonstrate LN metasurfaces with controllable SHG properties. Distinct enhancements for the SHG efficiency are observed at Mie-resonances. And by changing the geometric parameters thus the resonances of the metasurfaces, we manage to selectively boost the SHG efficiency at different wavelengths. Our results would pave a way for developing with high flexibility the novel compact nonlinear light sources for applications, such as biosensing and optical communications.
\end{abstract}

\keywords{Lithium niobate, Metasurface, Second harmonic generation, Resonance}
\maketitle

Second harmonic generation (SHG) plays a key role in extending spectral coverage of laser sources to wavelengths that are difficult to access by standard laser gain media. Such effect holds profound impacts on multiple applications including biological imaging microscopy, displays and communications.\cite{shen1984principles,boyd2003nonlinear} Because the intrinsic nonlinearities of materials are weak, the SHG efficiency is traditionally improved by increasing either the material lengths or pump intensities. And careful phase matching between fundamental and the second harmonic (SH) waves is further required to guarantee constructive accumulation of nonlinear waves along the macroscopic sized materials. Nowadays, as we are entering the nano era, realizing nanoscaled nonlinear optical devices has become a prominent research goal. Obviously, the adoption of the bulk media in the small devices will become impractical, in the mean time the above phase matching strategy would hardly play a role within subwavelength dimensions. Alternatively, artificial optical resonant nanostructures have been adopted to confine electromagnetic energy within nanometer scales to boost light-matter nonlinear interactions for improving the SHG efficiency.

As a typical kind of artificially nanostructured materials, optical metasurfaces which consist of subwavelength meta-atoms, provide us with a revolutionary concept to engineer nonlinear responses from the nanoscales.\cite{zheludev2012metamaterials,RenAM2020,wang2013double} The meta-atoms act as optical resonators and squeeze the electromagnetic energy into spaces even beyond the diffraction limit. Such tight field concentration strongly promotes the nonlinear effects. Furthermore, by manipulating the individual design and lattice arrangement of the meta-atoms, unprecedented freedom is obtained to control at will the polarization, spectrum, direction, phase or radiation profiles of the nonlinear electromagnetic radiations in both near- and far-fields. This benefits manufacturing with high flexibility the ultracompact photonic devices with multiple nonlinear optical functionalities.\cite{li2017nonlinear,krasnok2018nonlinear} And the distinguished low profile feature of the metasurfaces enables integration of multiple nonlinear optical functionalities onto a single ultra-thin photonic chip. 

Metallic plasmonic resonance was among the first exploited mechanism for the nonlinear metasurfaces. Employing the second-order nonlinearity originated from broken symmetry at the interfaces where the light fields are  confined tightly, enhanced SHG has been observed.\cite{klein2006second,tang2011nonlinear,rahmani2017nonlinear} But the high ohmic loss of metals has limited their practical use. In recent years, dielectric metasurfaces have emerged as promising alternatives to the metallic nanostructures. III-V semiconductors with non-centrosymmetric crystal structures and intrinsic second-order nonlinear responses, such as gallium arsenide (GaAs),\cite{LiuNL2016,vabishchevich2018enhanced,SautterNL2019} aluminum gallium arsenide (AlGaAs),\cite{carletti2018nonlinear,marino2019zero} and gallium phosphide (GaP)\cite{cambiasso2017bridging} have become the widely used constitutive materials. Benefiting from the high-quality Mie-resonances supported by their high refractive indices, the nonlinear interactions hence the SHG efficiencies are boosted efficiently. However, their available SHG spectral ranges are limited as a result of the relatively narrow band gap (about 1.4 to 1.5~eV): fundamental wave (FW) can hardly enter the visible range, and the SH with frequency above the band gap is absorbed resulting in suppression of the nonlinear efficiency. Moreover, growing crystalline III-V semiconductors on transparent substrates remains challenging because of the poor lattice match at the dielectric-semiconductor interfaces, therefore it is still rather challenging to engineer the SH properties in transmission direction through the substrate of these metasurfaces.\cite{Rahmani2018nonlinear} And due to the off-diagonal nature of the second order susceptibility tensor of these materials, the SH emission normal to the metasurface is usually prohibited, which imposes further constraints on their applicability.\cite{SautterNL2019}

\begin{figure} [pth]
\includegraphics[width=80mm]{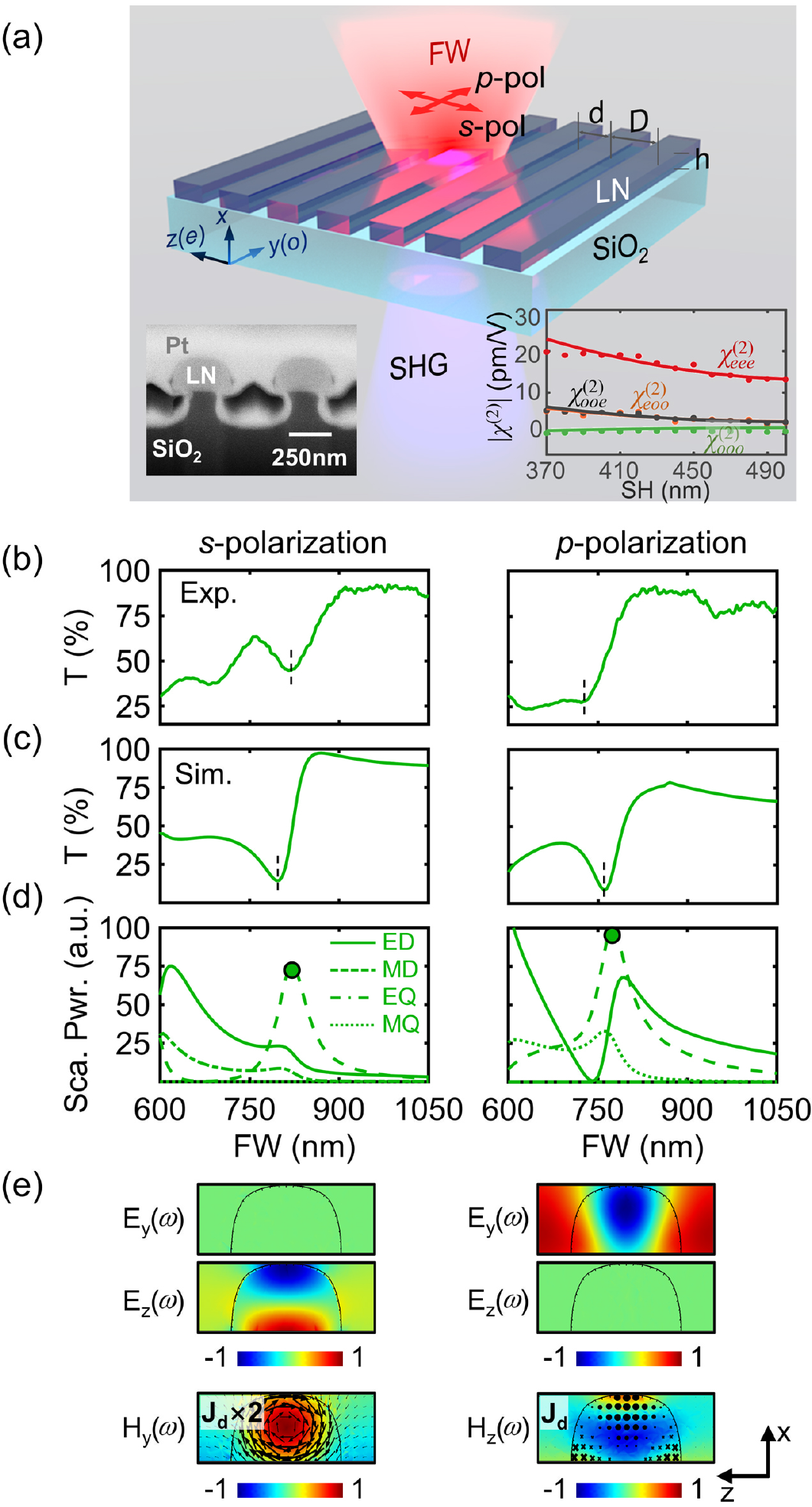} \caption{\label{fig1}
\textbf{Schematic of LN nonlinear metasurfaces.} \textbf{(a)} The metasurfaces are fabricated on a $x$-cut LN film residing on a fused quartz substrate. The LN principal crystallographic coordinate is used as the experimental coordinate system. Geometries of the metasurfaces are defined by ridge width $d$, period $D$, and height $h$. The grating ridges are oriented orthogonally to the optical axis of the LN ($z(e)$). Linearly polarized FW excites the metasurface from top, and the SH is collected on the opposite side. The light polarization perpendicular or parallel to the ridge direction is denoted as $s$- or $p$-polarization, respectively. Left inset gives a typical SEM image of cross section of the metasurface with $D$=600~nm. Right inset presents the measured second-order susceptibility of the LN film used in this study. Dots are experimental data and solid lines are eye-guides. \textbf{(b,c)} Experimentally measured and numerically simulated linear transmission spectra for array with $D$ of 600~nm for orthogonally polarized incidence. \textbf{(d)} Calculated linear scattering power spectra contributed by electric dipole (ED, solid line), magnetic dipole (MD, dashed line), electric quadrupole (EQ, dash-dotted line), and magnetic quadrupole (MQ, dotted line) under differently polarized excitations. \textbf{(e)} Field maps of the $D$ = 600~nm array at wavelengths marked by dots in (d). For $s$-polarization, displacement current density $\mathbf{J_d}$ are presented by black arrows. For $p$-polarization, crosses represent the $\mathbf{J_d}$ along $+y$ direction, and dots mean along $-y$ direction.}
\end{figure}

Thus to find a material for constructing the SHG nonlinear metasurfaces that can overcome above constraints has become an urgent problem to address. Lithium niobate (LN) can be an excellent solution. Thanks to its large band gap of 4~eV and outstanding second-order nonlinearities, the LN allows the SHG in a wide spectral range from ultraviolet to mid-infrared.\cite{Weis1985,Kongyongfa2020} Particularly, a breakthrough has been made recently in fabricating and commercializing thin-film LN on insulator (LNOI).\cite{levy1998fabrication,rabiei2004optical} The large refractive index contrast between the LN film and the transparent substrate (such as silica) leads to a well optical mode confinement and high optical intensity within the nanometers-thin LN layer, which makes LNOI very attractive and competitive for nonlinear integrated photonics applications.\cite{poberaj2012lithium,boes2018status,Qi2020Integrated} And the SHG from the LNOI micro- and nano-structures have been a subject under continuous researches in recent years such as LN nanowires,\cite{Sergeyev2013} photonic crystals,\cite{GeissAPL2010,Lu2012,Liang2017,Wei2018} waveguides,\cite{wang2017metasurface,Jankowski2020} hybrid LN-plasmonic nanopillars,\cite{Lehr2015} and micro-ring\cite{wang2019monolithic} or micro-disks.\cite{WangOL2018,kim2018anapole,PhysRevLettLin2019,Ye2020} On the other hand, researches of the LN metasurfaces are still in its infancy. Despite some theoretical studies of the SHG from the LN metasurfaces have been carried out,\cite{Fedotova2019Second,carletti2019second,li2020optical,carletti2020lithium} experimental realization still remains blank because of several obstacles in fabricating the LN metasurfaces, e.g.: lattice damages and ion-contamination are inevitably introduced to the LN in dry etching such as ion beam etching\cite{jiao2007ion,benchabane2009highly} or focussed ion beam milling (FIB)\cite{geiss2014photonic}; and the redeposition is serious during the fabrication because the meta-atoms are too compactly arrayed with subwavelength interval. All of these situations lead to an optical lossy amorphous layer at the surface of the meta-atoms which suppresses the optical resonance and the second order nonlinearities. Only until recently, the barriers in fabricating the LN metasurfaces have been removed using the FIB with great carefully adjusted processing parameters.\cite{Gao2019Lithium,fang2020second} And efficient tuning over the spectral resonance through varying the structural parameters has been demonstrated, which makes it possible to resonantly control the SHG from the LN metasurfaces.

In this paper, we demonstrate both numerically and experimentally the SHG from the LN nanograting metasurfaces over a broad spectral range. Both electric- and magnetic-natured Mie-resonances are steadily supported in the nanostructures, by which the SHG is observed to be resonantly enhanced. We manage to selectively boost the SHG efficiency at different wavelengths by tuning the geometric parameters and thus the resonances of the metasurfaces. The SHG efficiency as large as about 2$\times$10$^{-6}$ is achieved under a FW intensity of 2.05~GW/cm$^2$, which is about two times larger than that from the unstructured LN film. It is expected that apart from the SHG, our LN metasurfaces could be extended to engineer other frequency conversion nonlinearities, such as sum-frequency generation, four-wave mixing, parametric down conversion, etc., which can find a wide range of applications in communications, and quantum optics, et al.

A schematic of the SHG from the nonlinear LN metasurface is shown in Fig.~1. The metasurface consists of an array of LN nanogratings residing on a fused quartz substrate. The nanograting nanostructure design is adopted here not only because it is easy to fabricate, but also it can support Mie-resonances for local field enhancement\cite{Gao2019Lithium} which is essential to boost the nonlinear optical interactions. The LN ridge width is presented by $d$ and grating period by $D$. Height $h$ is 235~nm which is determined by thickness of the LN film (LNOI by NANOLN corporation) used for the metasurface fabrication. In order to utilize the largest second-order susceptibility ($\chi^{(2)}_{eee}$) of the LN, an $x$-cut LNOI wafer was used whose crystallographic optical axis ($z$-axis in Fig.~1(a)) located within the film plane. We fabricated the metasurfaces using a recently published procedure that involves FIB (Ga$^+$, 30~kV, 24~pA).\cite{Gao2019Lithium} And entire footprint of each metasurface array is 27$\times$27$~\mu$m$^2$. The fabricated metasurfaces are further etched by dilute HF to partially remove the SiO$_2$ under the LN ridges. Thus the refractive index contrast between the LN ridges and ambient is increased, leading to the improved electromagnetic field confinement inside the LN layer. Due to the angle divergence and scattering of the Ga$^+$ ion beams, cross sections of the fabricated ridges deviate the ideal rectangular design and are round shaped on the top. Inset of Fig.~1 gives a typical scanning electron microscope (SEM) image of the cross section of the fabricated metasurface with $D$=600~nm. The FW illuminates the metasurfaces from the LN side along the -$x$ direction. And the SHG is collected in the transmission direction. The FW polarization perpendicular or parallel to the ridge direction is denoted as $s$- or $p$-polarization, respectively. 

First of all we study the linear optical responses of the metasurfaces in visible and near-infrared wavelength range. And Fig.~1(b-e) exemplifies the results of the metasurface with $D$=600~nm. Left and right columns present the results for $s$- and $p$-polarized incidence, respectively. Figure~1(b) presents experimental transmission spectra measured by a commercial microscopic-spectrometer (IdeaOptics Technologies). We can identify a distinct resonance dip at about 820~nm for the $s$-polarized incidence and 730~nm for the $p$-polarization. And transmission plateaus are observed at longer wavelength for both polarizations. To compare our measured spectra with theory, we performed numerical simulations using the finite element method (COMSOL Multiphysics software). The LN ridges' geometric profiles in our models were extracted from the FIB cross section images of the fabricated samples (inset of Fig.~1(a)). The LN optical constants were taken from ellipsometric measurements. And the refractive index of the fused quartz underneath was set as 1.45. Periodic boundary conditions were used. The simulation results are shown in Fig.~1(c), which match well with the experimental ones particularly for the resonance wavelength positions. However, the experimental resonance dips are shallower with lower contrasts compared with the simulated results, which implies that the real samples have smaller resonance quality factors (Q-factors) than the numerical models. Such discrepancies may result from the Ga$^+$ contamination and the lattice damage during the FIB milling,\cite{geiss2014photonic} all of which would deteriorate the optical performance of the LN, but were not considered in our simulations.

\begin{figure} [tph]
\includegraphics[width=85mm]{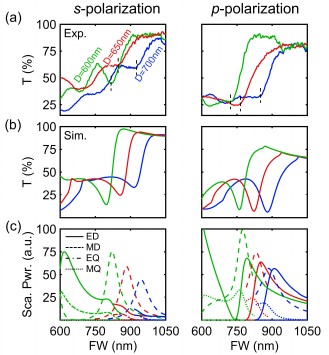} \caption{\label{fig2}
\textbf{Linear spectral properties of different LN nanograting metasurfaces.} \textbf{(a)} Experimentally measured transmission spectra for the metasurfaces with different $D$ for orthogonal polarizations. Green, red and blue lines correspond to $D$ of 600, 650 and 700~nm. \textbf{(b)} Simulated transmission spectra. \textbf{(c)} Multipolar decomposition of scattering power spectra.}
\end{figure}

In order to understand the nature of the spectral resonances, we performed cylindrical multipolar decomposition inside the nanogratings.\cite{chen2019singularities,chen2020global} And the calculated scattering power from different multipolar moments are shown in Fig.~1(d). We only retain the dipole and quadrupole components for clarity, and other higher-order components are negligible. It is clear that for the $s$-polarized incidence, the transmission dip correspond to a pronounced excitation of magnetic dipole (MD, dashed line). The near-field distributions on the ridge cross section at the MD resonance wavelength are plotted in left panel of Fig.~1(e). The resonant responses are characterized by the local field hot-spots (color maps) and displacement currents $\mathbf{J_d}$ (arrows) formed within the grating layer. The ridges are infinite in length, thus as a result of translational symmetry restriction, the electric field $E_y$ is zero under the $s$-polarization incidence. In the meanwhile, $E_z$ shows two hot spots confined in the close vicinity to the top and the bottom interfaces. And they are oscillating out-of-phase leading to an almost canceled effective electric dipole (ED) moment in the ridge, as indicated by the solid line in the left panel of Fig.~1(d). Circulating displacement currents are formed on the cross-section plane, which generates a hot spot in $H_y$ field and a strong MD dipole along the ridge direction. Such MD gives efficient multipolar scattering in the transmission direction leading to a pronounced MD peak. On the other hand, for the $p$-polarized incidence, the transmission resonance includes the overlapped contributions from ED, MD, and magnetic quadrupole (MQ). The electric hot-spots mainly localize near the top of the ridge and bottom of the air gaps (right panel of Fig.~1(e)). Moreover, resulting again from the translational symmetry restriction, the $\mathbf{J_d}$ only has the $y$-component. Despite the currents show no circulating features, however antiparallel currents are formed between the top and the bottom of LN ridge, hence the effective magnetic responses can be excited following the Ampere's circuital law (dashed and dotted lines in right panel of Fig.~1(d)).  

The employment of resonant nanostructures has an advantage of flexible spectral tunability. Considering that the LN shows a low optical dispersion in the studied wavelength range, thus by exploiting the scalability of Maxwell`s equations, the resonance wavelength could be easily tuned by scaling the metasurface geometries. For example, by changing the $D$ from 600~nm to 700~nm while fixing its ratio to the $d$ as 8: 5, the spectra could be efficiently red shifted, as shown in Fig.~2. It is worth noticing that the transmission resonances become sharper as $D$ decreases (both for experimental and simulated results in Fig.~2(a,b)). Such tendency is consistent with the stronger and narrowed multipolar resonances for smaller $D$ shown in Fig.~2(c). This implies that the larger local field magnitudes and better enhanced light-matter interactions for nanostructures with smaller $D$.

\begin{figure} [pth]
\includegraphics[width=85mm]{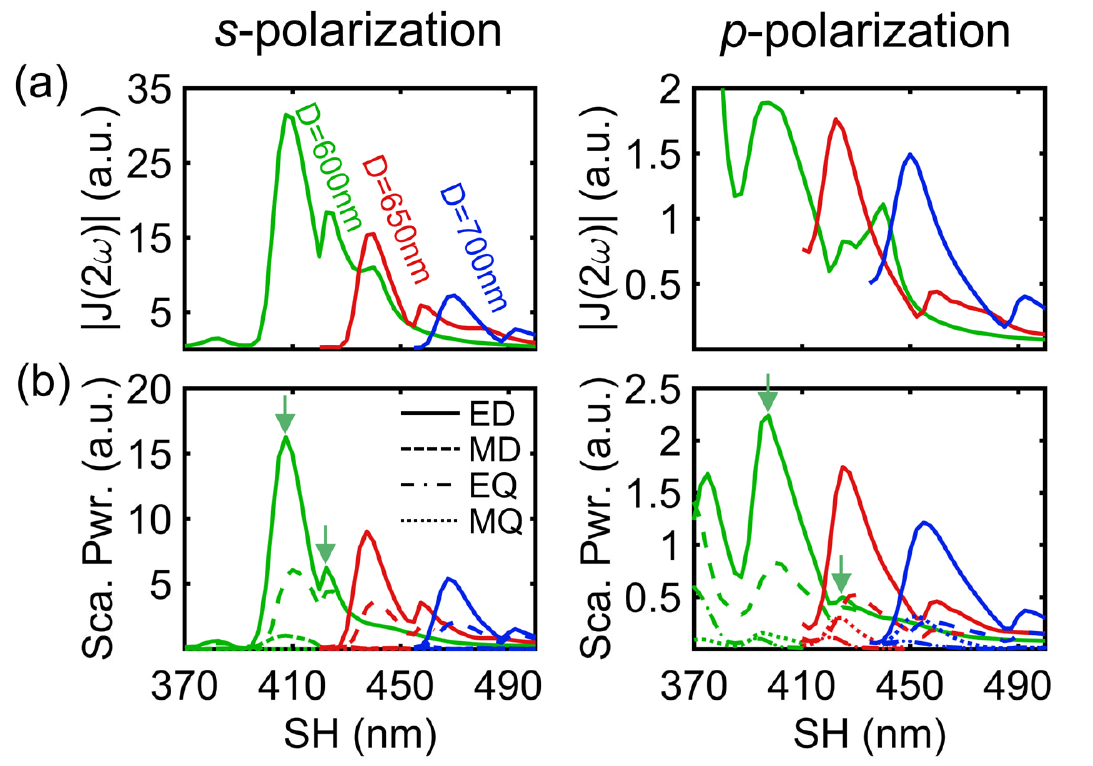} \caption{\label{fig3}
\textbf{Nonlinear responses from LN metasurfaces.} \textbf{(a)} Numerically simulated nonlinear displacement current density $|\mathbf{J}(2\omega)|$ excited inside the different metasurfaces. \textbf{(b)} Multipolar decomposition of scattering power by the nonlinear current $\mathrm{J}(2\omega)$.}
\end{figure}

The SHG relates to the second-order susceptibility $\mathbf{\chi^{(2)}}$ of the materials. Under strong enough FW excitation, the nonlinear displacement currents $\mathbf{J}(2\omega)$ would be generated inside the LN following $\mathbf{J}(2\omega)=4i\omega\epsilon_0\mathbf{\chi^{(2)}}:\mathbf{E}(\omega)\mathbf{E}(\omega)$. The $\mathbf{J}(2\omega)$ oscillates at twice the frequency of the FW fields ($\mathbf{E}(\omega)$) and acts as the secondary source radiating the SH to the far field. The LN crystal belongs to point group $3m$. Resulting from the symmetry restriction, the $\mathbf{\chi^{(2)}}$ has 11 nonvanishing components, in which only 4 elements are independent: $\chi^{(2)}_{ooe}$, $\chi^{(2)}_{eoo}$, $\chi^{(2)}_{ooo}$, and $\chi^{(2)}_{eee}$. The values of $\chi^{(2)}_{ijk}$ element of the LN (for SH between 370 and 500~nm) used in the simulations are measured based on the methods reported in Ref.[\onlinecite{Majj2020}] (given in the inset of Fig.~1(a)). We used the coupled electromagnetic waves frequency domain interfaces of COMSOL to numerically study the SHG properties of the LN metasurfaces.\cite{Majj2020,COMSOL} The incident fundamental pump intensity was set as 2.05 ~GW/cm$^2$ which was the same as our nonlinear experiments introduced in below. The spectra of the nonlinear current density $|\mathbf{J}(2\omega)|$ in different LN metasurfaces are shown in Fig.~3(a). Distinct spectral resonances are observed in the $|\mathbf{J}(2\omega)|$ spectra around the wavelengths corresponding to the FW resonances given in Fig.~2. To better gain the physical insight into the SHG from the metasurfaces, we further performed the multipolar decomposition using the $\mathbf{J}(2\omega)$ induced inside the LN, and the spectra are shown in Fig.~3(b). It is obvious that the ED is shown to dominate in all situations. And the nonlinear multipoles excited under the $s$-FW polarization are about one order of magnitude larger than the $p$-counterpart, which implies stronger nonlinear interactions under $s$-polarized FW excitation. Furthermore, two peaks are observed in each spectrum here (as indicated by arrows), which is in a distinct contrast to the single peak structure appears in the linear cases shown in Fig.~2(c). 

\begin{figure} [pth]
\includegraphics[width=85mm]{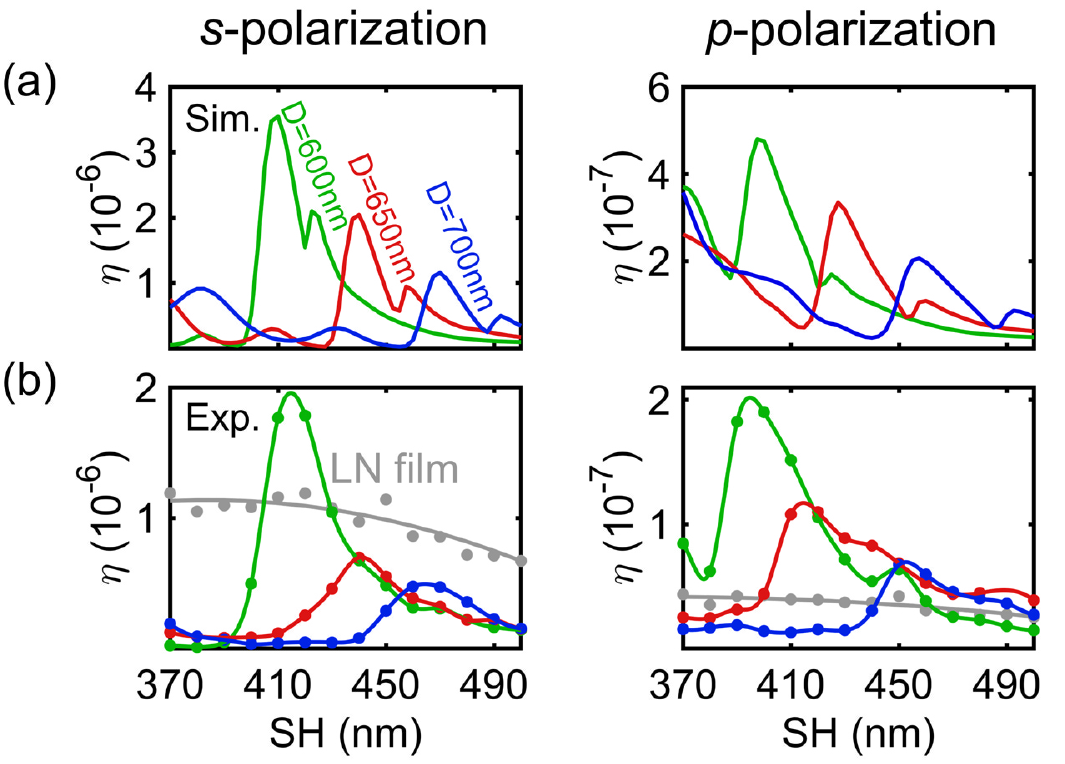} \caption{\label{fig4}
\textbf{Spectral dependence of SHG efficiencies $\eta$ from metasurfaces.} \textbf{(a)} Simulated SHG efficiencies from the different metasurfaces for orthogonal FW polarizations. \textbf{(b)} Experimentally measured SHG efficiencies for the different metasurfaces and unstructured LN film. Dots are experimental data and solid lines are eye-guides.} 
\end{figure}

We simulated the SHG power by integrating the Poynting vector over the output plane on the transmission side in the COMOSL models, which is further divided by the incident FW power to calculate the SHG conversion efficiency $\eta=P_{\mathrm{SHG}}/P_{\mathrm{FW}}$, as shown by solid lines in Fig.~4(a). The $\eta$ curves show a strong dependence on the wavelength, and present prominent enhancements around the wavelengths that correspond to the linear Mie resonances. Analogous to the spectral shifting in the linear transmission, the $\eta$ peaks are also shown to efficiently red-shift by increasing the $D$. Similar to the two peak structure in the nonlinear current density and multipolar spectra in Fig.~3, the $\eta$ spectra also show two peaks. The $\eta$ under the $s$-polarized pump is about one order of magnitude stronger than that of the $p$-excitation. This is because the electric near fields are polarized parallel to the optical axis of the LN crystal under the $s$-pump (shown in Fig.~1(e)), while $E$ fields are orthogonal to the optical axis for $p$-pump, thus the $\chi^{(2)}_{eee}$ (which is more than 3 times larger than other $\chi^{(2)}$ components as shown in Fig.~1(a)) is more efficiently utilized by the $s$-pump. And the $\eta$ achieves about 3.5$\times 10^{-6}$ and peaks at 410~nm for $D$= 600~nm under $s$-polarized FW.

To experimentally characterize the SHG from the metasurfaces, the samples were illuminated by a tunable femtosecond laser (Mai Tai, Spectra-Physics, 200 fs, 80 MHz). The pump laser beam was focused onto the LN metasurfaces by a 10$\times$ objective (N.A.=0.25) forming a spot with a radius of 4.7~$\mu$m. The SH signal was collected in the transmission side using an UV-objective (Thorlabs, 20$\times$, N.A.=0.38) whereas the transmitted pump wave was filtered-out using a short-pass filter (BG40 colored glass). We measured the SHG signal by sweeping the pump wavelength across a range from 740 to 1000~nm in a of step of 20~nm and recording the SHG intensity spanning from 370 to 500~nm. The incident average pump power was kept constant at around 23~mW corresponding to a peak irradiance of about 2.05~GW/cm$^2$. And the SH power was evaluated using a calibrated spectrometer. As shown in Fig.~4(b), the experimental $\eta$ curves clearly show resonant peaks, whose wavelength positions are consistent with the simulated ones and also show red-shifts for the larger $D$. It is worth noticing that the larger SHG is achieved for smaller $D$. Such phenomenon is consistent with the stronger linear multipole resonances for the smaller $D$ shown in Fig.~1(d), which implies that the LN nanostructures interact more efficiently with the incident FW field. The maximum $\eta$ in our experiment achieves about 2$\times10^{-6}$ from the $D$=600~nm metasurface for $s$-polarized FW incidence, which is about two times larger than that of the unstructured LN film (grey line). Furthermore, consistent with the numerical prediction, the $\eta$ under $p$-FW polarization is about one order of magnitude smaller than that of the $s$-counterpart, while up to 5 times enhancement in the $\eta$ is achieved for the $D$=600~nm metasurface around the SH wavelength of 400~nm. The experimental $\eta$ curves show smaller magnitudes and broader spectral widths than the simulated ones. Such differences could be explained by the much lower Q-factors and thus less nonlinear interaction strengths in the real samples than the simulated cases. Furthermore, imperfections in the alignment of our experimental setup may also reduce the measured $\eta$.

In conclusion, we have experimentally demonstrated the nonlinear LN metasurfaces with controllable SHG properties. Both electric- and magnetic-natured Mie-resonances are steadily supported in the LN metasurfaces, allowing for the resonantly enhanced nonlinear optical interactions. The SHG efficiency is observed to be boosted efficiently at the resonance wavelengths and achieves about 2$\times10^{-6}$ at the 2.05~GW/cm$^2$ pump intensity, which is about two times larger than that of the unstructured LN film. Furthermore, we manage to tune the SHG efficiency spectra by changing the geometric parameters thus the Mie-resonances of the LN metasurfaces. The experimental results are consistent with the numerical simulations. Our nonlinear LN metasurfaces would benefit manufacturing with high flexibility the ultracompact photonic devices showing novel nonlinear optical functionalities. And our results can also be applied for other nonlinear processes such as four-wave mixing, sum frequency generation, parametric down conversion, et al, which may find wide applications in biosensing, quantum optical communications, display, and so on. 

\begin{acknowledgments}
This work was supported by National Key R\&D Program of China (2017YFA0305100, 2017YFA0303800); National Natural Science Foundation of China (61775106, 11904182, 11711530205, 11374006, 11774185, 91750204); the 111 Project (B07013); PCSIRT (IRT0149); the Open Research Program of Key Laboratory of 3D Micro/Nano Fabrication and Characterization of Zhejiang Province; the Fundamental Research Funds for the Central Universities (010-63201008); the Tianjin youth talent support program. We thank the Nanofabrication Platform of Nankai University for fabricating samples.
\end{acknowledgments}


\begin{thebibliography}{10}
\newcommand{\enquote}[1]{``#1''}

\bibitem{shen1984principles}
Y.-R. Shen, \emph{The principles of nonlinear optics} (Wiley-Interscience,
  1984).

\bibitem{boyd2003nonlinear}
R.~W. Boyd, \emph{Nonlinear Optics} (Elsevier, 2003).

\bibitem{zheludev2012metamaterials}
N.~I. Zheludev and Y.~S. Kivshar, \enquote{From metamaterials to metadevices,}
  Nature Mater. \textbf{11}, 917--924 (2012).

\bibitem{RenAM2020}
M.~Ren, W.~Cai, and J.~Xu, \enquote{Tailorable dynamics in nonlinear optical
  metasurfaces,} Adv. Mater. \textbf{32}, 1806317 (2020).

\bibitem{wang2013double}
J.~Wang, C.~Fan, J.~He, P.~Ding, E.~Liang, and Q.~Xue, \enquote{Double fano
  resonances due to interplay of electric and magnetic plasmon modes in planar
  plasmonic structure with high sensing sensitivity,} Opt. express \textbf{21},
  2236--2244 (2013).

\bibitem{li2017nonlinear}
G.~Li, S.~Zhang, and T.~Zentgraf, \enquote{Nonlinear photonic metasurfaces,}
  Nature Rev. Mater. \textbf{2}, 1--14 (2017).

\bibitem{krasnok2018nonlinear}
A.~Krasnok, M.~Tymchenko, and A.~Al{\`u}, \enquote{Nonlinear metasurfaces: a
  paradigm shift in nonlinear optics,} Mater. Today \textbf{21}, 8--21 (2018).

\bibitem{klein2006second}
M.~W. Klein, C.~Enkrich, M.~Wegener, and S.~Linden, \enquote{Second-harmonic
  generation from magnetic metamaterials,} Science \textbf{313}, 502--504
  (2006).

\bibitem{tang2011nonlinear}
S.~Tang, D.~J. Cho, H.~Xu, W.~Wu, Y.~R. Shen, and L.~Zhou, \enquote{Nonlinear
  responses in optical metamaterials: theory and experiment,} Opt. Express
  \textbf{19}, 18283--18293 (2011).

\bibitem{rahmani2017nonlinear}
M.~Rahmani, A.~S. Shorokhov, B.~Hopkins, A.~E. Miroshnichenko, M.~R.
  Shcherbakov, R.~Camacho-Morales, A.~A. Fedyanin, D.~N. Neshev, and Y.~S.
  Kivshar, \enquote{Nonlinear symmetry breaking in symmetric oligomers,} ACS
  Photonics \textbf{4}, 454--461 (2017).

\bibitem{LiuNL2016}
S.~Liu, M.~B. Sinclair, S.~Saravi, G.~A. Keeler, Y.~Yang, J.~Reno, G.~M. Peake,
  F.~Setzpfandt, I.~Staude, T.~Pertsch, and I.~Brener, \enquote{Resonantly
  enhanced second-harmonic generation using iii-v semiconductor all-dielectric
  metasurfaces,} Nano Lett. \textbf{16}, 5426--5432 (2016).

\bibitem{vabishchevich2018enhanced}
P.~P. Vabishchevich, S.~Liu, M.~B. Sinclair, G.~A. Keeler, G.~M. Peake, and
  I.~Brener, \enquote{Enhanced second-harmonic generation using broken symmetry
  iii--v semiconductor fano metasurfaces,} ACS Photonics \textbf{5}, 1685--1690
  (2018).

\bibitem{SautterNL2019}
J.~D. Sautter, L.~Xu, A.~E. Miroshnichenko, M.~Lysevych, I.~Volkovskaya, D.~A.
  Smirnova, R.~Camacho-Morales, K.~Zangeneh~Kamali, F.~Karouta, K.~Vora, H.~H.
  Tan, M.~Kauranen, I.~Staude, C.~Jagadish, D.~N. Neshev, and M.~Rahmani,
  \enquote{Tailoring second-harmonic emission from (111)-gaas nanoantennas,}
  Nano Lett. \textbf{19}, 3905--3911 (2019).

\bibitem{carletti2018nonlinear}
L.~Carletti, G.~Marino, L.~Ghirardini, V.~F. Gili, D.~Rocco, I.~Favero,
  A.~Locatelli, A.~V. Zayats, M.~Celebrano, M.~Finazzi \emph{et~al.},
  \enquote{Nonlinear goniometry by second-harmonic generation in algaas
  nanoantennas,} ACS Photonics \textbf{5}, 4386--4392 (2018).

\bibitem{marino2019zero}
G.~Marino, C.~Gigli, D.~Rocco, A.~Lema{\^\i}tre, I.~Favero, C.~De~Angelis, and
  G.~Leo, \enquote{Zero-order second harmonic generation from
  algaas-on-insulator metasurfaces,} ACS Photonics \textbf{6}, 1226--1231
  (2019).

\bibitem{cambiasso2017bridging}
J.~Cambiasso, G.~Grinblat, Y.~Li, A.~Rakovich, E.~Cort{\'e}s, and S.~A. Maier,
  \enquote{Bridging the gap between dielectric nanophotonics and the visible
  regime with effectively lossless gallium phosphide antennas,} Nano Lett.
  \textbf{17}, 1219--1225 (2017).

\bibitem{Rahmani2018nonlinear}
M.~Rahmani, G.~Leo, I.~Brener, A.~V. Zayats, S.~A. Maier, C.~D. Angelis,
  H.~Tan, V.~F. Gili, F.~Karouta, R.~Oulton, K.~Vora, M.~Lysevych, I.~Staude,
  L.~Xu, A.~E. Miroshnichenko, C.~Jagadish, and D.~N. Neshev,
  \enquote{Nonlinear frequency conversion in optical nanoantennas and
  metasurfaces: materials evolution and fabrication,} Opto-Electron. Adv.
  \textbf{1}, 180021 (2018).

\bibitem{Weis1985}
R.~S. Weis and T.~K. Gaylord, \enquote{Lithium niobate: Summary of physical
  properties and crystal structure,} Appl. Phys. A \textbf{37}, 191--203
  (1985).

\bibitem{Kongyongfa2020}
Y.~Kong, F.~Bo, W.~Wang, D.~Zheng, H.~Liu, G.~Zhang, R.~Rupp, and J.~Xu,
  \enquote{Recent progress in lithium niobate: Optical damage, defect
  simulation, and on-chip devices,} Adv. Mater. \textbf{32}, 1806452 (2020).

\bibitem{levy1998fabrication}
M.~Levy, R.~Osgood~Jr, R.~Liu, L.~Cross, G.~Cargill~III, A.~Kumar, and
  H.~Bakhru, \enquote{Fabrication of single-crystal lithium niobate films by
  crystal ion slicing,} Appl. Phys. Lett. \textbf{73}, 2293--2295 (1998).

\bibitem{rabiei2004optical}
P.~Rabiei and P.~Gunter, \enquote{Optical and electro-optical properties of
  submicrometer lithium niobate slab waveguides prepared by crystal ion slicing
  and wafer bonding,} Appl. Phys. Lett. \textbf{85}, 4603--4605 (2004).

\bibitem{poberaj2012lithium}
G.~Poberaj, H.~Hu, W.~Sohler, and P.~Guenter, \enquote{Lithium niobate on
  insulator (lnoi) for micro-photonic devices,} Laser \& Photonics Rev.
  \textbf{6}, 488--503 (2012).

\bibitem{boes2018status}
A.~Boes, B.~Corcoran, L.~Chang, J.~Bowers, and A.~Mitchell, \enquote{Status and
  potential of lithium niobate on insulator (lnoi) for photonic integrated
  circuits,} Laser \& Photonics Rev. \textbf{12}, 1700256 (2018).

\bibitem{Qi2020Integrated}
Y.~Qi and Y.~Li, \enquote{Integrated lithium niobate photonics,} Nanophotonics
  \textbf{9}, 1287 -- 1320 (2020).

\bibitem{Sergeyev2013}
A.~Sergeyev, R.~Geiss, A.~S. Solntsev, A.~Steinbr\"{u}ck, F.~Schrempel, E.-B.
  Kley, T.~Pertsch, and R.~Grange, \enquote{Second-harmonic generation in
  lithium niobate nanowires for local fluorescence excitation,} Opt. Express
  \textbf{21}, 19012--19021 (2013).

\bibitem{GeissAPL2010}
R.~Geiss, S.~Diziain, R.~Iliew, C.~Etrich, H.~Hartung, N.~Janunts,
  F.~Schrempel, F.~Lederer, T.~Pertsch, and E.-B. Kley, \enquote{Light
  propagation in a free-standing lithium niobate photonic crystal waveguide,}
  Appl. Phys. Lett. \textbf{97}, 131109 (2010).

\bibitem{Lu2012}
H.~Lu, B.~Sadani, N.~Courjal, G.~Ulliac, N.~Smith, V.~Stenger, M.~Collet, F.~I.
  Baida, and M.-P. Bernal, \enquote{Enhanced electro-optical lithium niobate
  photonic crystal wire waveguide on a smart-cut thin film,} Opt. Express
  \textbf{20}, 2974--2981 (2012).

\bibitem{Liang2017}
H.~Liang, R.~Luo, Y.~He, H.~Jiang, and Q.~Lin, \enquote{High-quality lithium
  niobate photonic crystal nanocavities,} Optica \textbf{4}, 1251--1258 (2017).

\bibitem{Wei2018}
D.~Wei, C.~Wang, H.~Wang, X.~Hu, D.~Wei, X.~Fang, Y.~Zhang, D.~Wu, Y.~Hu,
  J.~Li, Z.~Shining, and X.~Min, \enquote{Experimental demonstration of a
  three-dimensional lithium niobate nonlinear photonic crystal,} Nat. Photonics
  \textbf{12}, 596--600 (2018).

\bibitem{wang2017metasurface}
C.~Wang, Z.~Li, M.-H. Kim, X.~Xiong, X.-F. Ren, G.-C. Guo, N.~Yu, and
  M.~Lon{\v{c}}ar, \enquote{Metasurface-assisted phase-matching-free second
  harmonic generation in lithium niobate waveguides,} Nat. Commun. \textbf{8},
  1--7 (2017).

\bibitem{Jankowski2020}
M.~Jankowski, C.~Langrock, B.~Desiatov, A.~Marandi, C.~Wang, M.~Zhang, C.~R.
  Phillips, M.~Lon\v{c}ar, and M.~M. Fejer, \enquote{Ultrabroadband nonlinear
  optics in nanophotonic periodically poled lithium niobate waveguides,} Optica
  \textbf{7}, 40--46 (2020).

\bibitem{Lehr2015}
D.~Lehr, J.~Reinhold, I.~Thiele, H.~Hartung, K.~Dietrich, C.~Menzel,
  T.~Pertsch, E.-B. Kley, and A.~T$\ddot{u}$nnermann, \enquote{Enhancing second
  harmonic generation in gold nanoring resonators filled with lithium niobate,}
  Nano Lett. \textbf{15}, 1025--1030 (2015).

\bibitem{wang2019monolithic}
C.~Wang, M.~Zhang, M.~Yu, R.~Zhu, H.~Hu, and M.~Loncar, \enquote{Monolithic
  lithium niobate photonic circuits for kerr frequency comb generation and
  modulation,} Nat. Commun. \textbf{10}, 1--6 (2019).

\bibitem{WangOL2018}
L.~Wang, C.~Wang, J.~Wang, F.~Bo, M.~Zhang, Q.~Gong, M.~Lon\v{c}ar, and Y.-F.
  Xiao, \enquote{High-q chaotic lithium niobate microdisk cavity,} Opt. Lett.
  \textbf{43}, 2917--2920 (2018).

\bibitem{kim2018anapole}
K.-H. Kim and W.-S. Rim, \enquote{Anapole resonances facilitated by high-index
  contrast between substrate and dielectric nanodisk enhance vacuum ultraviolet
  generation,} ACS Photonics \textbf{5}, 4769--4775 (2018).

\bibitem{PhysRevLettLin2019}
J.~Lin, N.~Yao, Z.~Hao, J.~Zhang, W.~Mao, M.~Wang, W.~Chu, R.~Wu, Z.~Fang,
  L.~Qiao, W.~Fang, F.~Bo, and Y.~Cheng, \enquote{Broadband quasi-phase-matched
  harmonic generation in an on-chip monocrystalline lithium niobate microdisk
  resonator,} Phys. Rev. Lett. \textbf{122}, 173903 (2019).

\bibitem{Ye2020}
X.~Ye, S.~Liu, Y.~Chen, Y.~Zheng, and X.~Chen, \enquote{Sum-frequency
  generation in lithium-niobate-on-insulator microdisk via modal phase
  matching,} Opt. Lett. \textbf{45}, 523--526 (2020).

\bibitem{Fedotova2019Second}
A.~Fedotova, M.~Younesi, J.~Sautter, M.~Steinert, R.~Geiss, T.~Pertsch,
  I.~Staude, and F.~Setzpfandt, \enquote{Second-harmonic generation in lithium
  niobate metasurfaces,} in \enquote{European Quantum Electronics Conference,}
  (2019), p. ef\_1\_2.

\bibitem{carletti2019second}
L.~Carletti, C.~Li, J.~Sautter, I.~Staude, C.~De~Angelis, T.~Li, and D.~N.
  Neshev, \enquote{Second harmonic generation in monolithic lithium niobate
  metasurfaces,} Opt. Express \textbf{27}, 33391--33398 (2019).

\bibitem{li2020optical}
Y.~Li, Z.~Huang, Z.~Sui, H.~Chen, X.~Zhang, W.~Huang, H.~Guan, W.~Qiu, J.~Dong,
  W.~Zhu \emph{et~al.}, \enquote{Optical anapole mode in nanostructured lithium
  niobate for enhancing second harmonic generation,} Nanophotonics \textbf{1}
  (2020).

\bibitem{carletti2020lithium}
L.~Carletti, A.~Zilli, F.~Moia, A.~Toma, M.~Finazzi, M.~Celebrano,
  C.~De~Angelis, and D.~Neshev, \enquote{Lithium niobate metasurfaces for
  second-harmonic generation,} in \enquote{2020 Conference on Lasers and
  Electro-Optics (CLEO),}  (IEEE, 2020), pp. 1--2.

\bibitem{jiao2007ion}
Y.~Jiao, K.-M. Wang, X.-L. Wang, L.~Wang, C.-L. Jia, Y.~Jiang, J.-H. Zhang, and
  F.~Lu, \enquote{Ion beam etched diffraction gratings in fused quartz and
  lithium niobate,} Surf. Coat. Technol. \textbf{201}, 5046--5049 (2007).

\bibitem{benchabane2009highly}
S.~Benchabane, L.~Robert, J.-Y. Rauch, A.~Khelif, and V.~Laude, \enquote{Highly
  selective electroplated nickel mask for lithium niobate dry etching,} J.
  Appl. Phys. \textbf{105}, 094109 (2009).

\bibitem{geiss2014photonic}
R.~Geiss, S.~Diziain, M.~Steinert, F.~Schrempel, E.-B. Kley, A.~T{\"u}nnermann,
  and T.~Pertsch, \enquote{Photonic crystals in lithium niobate by combining
  focussed ion beam writing and ion-beam enhanced etching,} Phys. Status Solidi
  A \textbf{211}, 2421--2425 (2014).

\bibitem{Gao2019Lithium}
B.~Gao, M.~Ren, W.~Wu, H.~Hu, W.~Cai, and J.~Xu, \enquote{Lithium niobate
  metasurfaces,} Laser Photonics Rev. \textbf{13}, 1800312 (2019).

\bibitem{fang2020second}
B.~Fang, H.~Li, S.~Zhu, and T.~Li, \enquote{Second-harmonic generation and
  manipulation in lithium niobate slab waveguides by grating metasurfaces,}
  Photonics Res. \textbf{8}, 1296--1300 (2020).

\bibitem{chen2019singularities}
W.~Chen, Y.~Chen, and W.~Liu, \enquote{Singularities and poincar{\'e} indices
  of electromagnetic multipoles,} Phys. Rev. Lett. \textbf{122}, 153907 (2019).

\bibitem{chen2020global}
W.~Chen, Q.~Yang, Y.~Chen, and W.~Liu, \enquote{Global mie scattering:
  Polarization morphologies and the underlying topological invariant,} ACS
  Omega  (2020).

\bibitem{Majj2020}
J.~Ma, J.~Chen, M.~Ren, W.~Wu, W.~Cai, and J.~Xu, \enquote{Second-harmonic
  generation and its nonlinear depolarization from lithium niobate thin films,}
  Opt. Lett. \textbf{45}, 145--148 (2020).

\bibitem{COMSOL}
\url{https://comsol.com/model/second-harmonic-generation-24151}.

\end{thebibliography}

\end{document}